\documentclass[onecolumn,prb]{revtex4-2}
\usepackage{graphicx,psfrag,color}
\usepackage{slashed,amssymb}
\def\bc{\begin{center}}
\def\ec{\end{center}}

\def\beq{\begin{equation}}
\def\eeq{\end{equation}}

\def\bq{{\bf q}}

\def\bc{{\bf c}}

\def\text{{\rm}} 

\begin{document}

\title{Pseudospin order of the Bogoliubov de Gennes Hamiltonian}
\author{Klaus Ziegler$^{1,2}$\\ 
$^1$Institut f\"ur Physik, Universit\"at Augsburg\\
D-86135 Augsburg, Germany\\ 
$^{2}$Physics Department, New York City College of Technology\\
 The City University of New York\\
Brooklyn, NY 11201, USA \\
}
\date{\today}


\begin{abstract}
We investigate a two-dimensional superconducting system with a
smoothly and periodically varying order parameter, which is modulated along
one direction while remaining uniform in the perpendicular direction, leading to a spatially
periodic superconducting phase. 
It is known that the phase of the order parameter determines the relative phase
of the quasiparticle components. Here we demonstrate that the latter becomes observable
through the associated Bloch vector, which provides a topological invariant that links the
Bloch vector to an integer winding number.
The Bloch vector can be understood as the expectation value of the pseudospin, 
for which the physical manifestation of the winding number is a helical ordering.
For the Bogoliubov–de Gennes Hamiltonian with an edge we have both plane-wave solutions
describing bulk states and exponentially localized solutions that correspond to edge modes.
The analytic bulk–edge connection is employed to identify the conditions under which the
edge states emerge from the bulk spectrum. 
The pseudospin ordering rotates under the energy change of the edge
modes but is energy-independent for the bulk modes, except for the opposite sign in the
upper and the lower band.
These results establish a direct connection between the spatial modulation of the order
parameter, the topological structure of the Bloch vector, 
as well as the relation between the order parameter phase and the pseudospin orientation
of the edge modes.
\end{abstract}

\maketitle

\section{Introduction}

Quasiparticles in a superconductor are described by the Bogoliubov de Gennes (BdG) Hamiltonian,
which represents a two-band model. There is a gap $2|\Delta|$ between these bands, which is determined 
by the density of the superconducting electrons through the magnitude of the superconducting order 
parameter $|\Delta|$. The order parameter itself has a phase factor, such that $\Delta=|\Delta|e^{i\varphi}$
describes the superconducting state. The model of the superconductor is invariant under a global
phase shift $\varphi\to\varphi+\varphi_0$. Thus, we can remove a uniform phase by such a shift
operation, which means that a uniform phase is not observable. A spatially non-uniform phase,
on the other hand, cannot be removed by a global phase shift. It turns out that it is observable and
related to a gauge field in which the electrons move. A typical example is an external magnetic field
(e.g., the Abrikosov lattice of a type II superconductor~\cite{abrikosov57}).
Other examples for a system with a spatially varying order-parameter phase are Josephson junctions~\cite{josephson62,blonder82,spuntarelli10}.
For the following, we are mostly interested in a periodically changing order-parameter phase, as
it appears in the Fulde-Ferrell-Larkin-Ovchinnikov (FFLO) 
superconductor~\cite{fulde64,larkin64,zheng13,guo17,zhang22,shimahara09}, 
in a rotating superfluid~\cite{tisza38,landau41}, in a moving polariton condensate~\cite{carusotto13} or in a sliding
charge-density wave~\cite{gruener88}. The one-dimensional Su-Schrieffer-Heeger (SSH) Hamiltonian
with long-range hopping is a more recent example of a two-band system with modulated order
parameter~\cite{liu23}.

For a given microscopic model of interacting electrons, the superconducting order parameter is usually 
determined by the self-consistent mean-field (BCS) approach~\cite{bardeen57,tinkham96}, while the
fluctuations out of the superconducting condensate of Cooper pairs are described by the BdG
Hamiltonian. In this article we will focus on the quasiparticles and assume that the order parameter
$\Delta$ is obtained from the BCS approach. The BdG Hamiltonian can be expanded in terms of
Pauli matrices as $H_{\rm BdG}=\vec{h}_{\rm BdG}\cdot\vec{\sigma}$, which belongs to a larger class of
$SU(2)$ Hamiltonians. 
In general, the $SU(2)$ Hamiltonian of the form $\vec{h}\cdot\vec{\sigma}$ has been used as a 
prototype for topological states in condensed matter
to describe a number of fascinating properties in special materials, ranging from 2D 
graphene-like materials (e.g., the Haldane model) to 3D Weyl 
semimetals~\cite{haldane88,yakovenko90,hatsugai93,kane05a,kane05b,bernevig06,
armitage18,burkov18,fradkin}. 
An interpretation of this Hamiltonian is that of a spin $\vec{\sigma}$ is coupled to a
(operator valued) pseudomagnetic field $\vec{h}$.
Many of the specific properties of this type of Hamiltonian and its eigenfunctions originate in the 
robust winding number of the spinor states. A deeper understanding of the connection between the $SU(2)$
Hamiltonian and the winding number of the spinor reveals robust physical properties of
systems that are governed by such a Hamiltonian. In particular, we will study in this article how
the phase of the superconducting order parameter affects the winding number and how this
can be described by a Bloch vector and a related topological invariant, which is an observable quantity.
In the discussion we distinguish between bulk and edge modes. 

\begin{figure}[t]
\begin{center}
\includegraphics[width=0.4\linewidth]{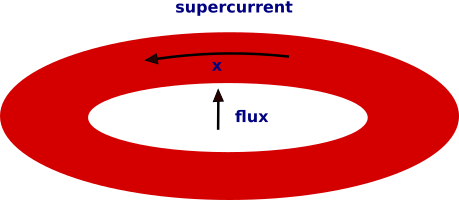}
\caption{
A macroscopic vortex on a superconducting ring with radius $R$: the 
supercurrent induces a $2\pi n$ periodic order parameter $\Delta(x)=|\Delta|\exp(ix/L)$ with $L=l/2\pi n$ for the
circumference $l=2\pi R$ of the ring,
where the integers $n=0,\pm1,\ldots$ are enforced by the periodic boundary conditions in $x$ direction. 
The specific value of $n$ depends on the strength of the supercurrent, where an increasing supercurrent 
yields an increasing quantized winding number for the quasiparticle states.}
\label{fig:1}
\end{center}
\end{figure}

The winding number is often associated with a robust behavior due to its connection
with topological invariants, for instance, in chiral quantum systems. Therefore, a weak
perturbation might be insufficient for a change but we must rely on a strong and macroscopic
intervention~\cite{aharonov59}, usually caused by a symmetry change. What determines the winding 
number in a given system 
and how can we change it? And is it possible to control it by an external and macroscopic method? 
To answer these questions, we consider a ring made of a superconducting material, as schematically 
presented in Fig. \ref{fig:1}, and apply a phase modulated order parameter. This can be understood as
a macroscopic vortex in the superconducting state~\cite{bruyndoncx99,kanda04,liu21}.
The phase modulation is determined by the boundary conditions of the system, since the order parameter
must be a continuous function. This leads to a quantization of the order 
parameter phase.  In this paper we will discuss how the latter affects the quasiparticle properties, which are 
subject to the BdG Hamiltonian.

Of special interest are edge modes, which appear in the presence of sample edges. Examples are
a finite disk or a cylinder of finite length.  In contrast to systems with periodic boundary conditions, where
many orthogonal bulk modes exist, there is typically only a single edge mode which obeys the boundary 
conditions due to its exponential behavior. This, however, can depend on other degrees of freedom. For
instance, on the coordinate perpendicular to the edge.
Starting from periodic bulk modes, we obtain the exponential
edge modes in a circular geometry by an analytic continuation of the wave number for the BdG 
Hamiltonian with real order parameter~\cite{liu25}
and for the Dirac-like Hamiltonians~\cite{ziegler25a,ziegler25b}.
This approach can be extended to other geometries and other Hamiltonians. In general, this requires the 
analytic continuation of several parameters. For instance, on a square we must apply the analytic 
continuation for both components of the wave vector. In the present case we will consider the BdG Hamiltonian
with complex order parameter and only edges in  one direction to
keep the calculations as simple as possible. In Fig. \ref{fig:1} we can create an edge by cutting the 
ring at one angle. Then the order parameter does not need to be periodic anymore.

The article is organized as follows. In Sect. \ref{sect:model} the BdG Hamiltonian, its energy dispersion
and its eigenspinors are defined and discussed. Then the Bloch vector, a topological invariant for the
winding number on the Bloch sphere and the order-parameter phase transformation are discussed in 
Sect. \ref{sect:bloch_vector}. After this general part, in Sect. \ref{sect:examples} we turn to two 
specific realizations of the BdG Hamiltonian: in Sect. \ref{sect:cont_bdg} we study it in the continuum and
in Sect. \ref{sect:bdg_tb} for the tight-binding realization on the square lattice. Edge modes and their relations
with the bulk modes are discussed in Sect. \ref{sect:edges}. Finally, in Sect. \ref{sect:discussion} the results
of the BdG Hamiltonian are summarized and compared with the $\pi$-flux Hamiltonian. 

\section{Model}
\label{sect:model}

A large class of
two-dimensional systems is described by the $SU(2)$ Hamiltonian $\vec{h}\cdot\vec{\sigma}$ 
with the Pauli vector $\vec{\sigma}=(\sigma_1,\sigma_2,\sigma_3)^T$ and with the three-dimensional 
vector~\cite{haldane88,yakovenko90,hatsugai93,kane05a,kane05b,bernevig06,armitage18,burkov18}
\beq
\label{su(2)_ham}
\vec{h}=(\Delta',\Delta'',D)^T
,
\eeq
where each vector component can be a self-adjoint differential operator.  For the following study, however,
we focus on the case where only $D$ is a 
differential or difference operator on a two-dimensional space with coordinates $(x,y)$, while $\Delta'$, $\Delta''$
are the real and imaginary parts of a periodic function $\Delta(x)$ in space. $\Delta=\Delta'+i\Delta''$
can be considered as a complex order parameter, as it appears in the BdG Hamiltonian.
Specific examples are discussed in Sect. \ref{sect:examples}. We assume that there is
a pair of complex numbers $\gamma_j$ ($j=1,2$) with 
$De^{\gamma_j x}=d(\gamma_j)e^{\gamma_j x}$.
In other words, $e^{\gamma_j x}$ is eigenfunction of $D$ with an eigenvalue
$d(\gamma_j)$. With $\Delta(x)=|\Delta|e^{ix/L}$ and a uniform $|\Delta|$ we introduce the spinor
\beq
\label{spinor0}
\Psi(x)
=\pmatrix{
a_1 e^{\gamma_1x}\cr
a_2e^{\gamma_2x}\cr
}
\eeq
to formulate the energy eigenvalue problem $\vec{h}\cdot\vec{\sigma}\Psi_x=E\Psi_x$ in $2\times2$ 
matrix notation as
\beq
\label{eigenvalue_problem}
\pmatrix{
d(\gamma_1)a_1e^{\gamma_1x}+|\Delta| a_2e^{(\gamma_2-i/L) x} \cr
|\Delta|a_1e^{(\gamma_1 +i/L)x}-d(\gamma_2)a_2e^{\gamma_2 x} \cr
}
=E\pmatrix{
a_1e^{\gamma_1 x} \cr
a_2e^{\gamma_2 x} \cr
}
\eeq
with a real eigenvalue $E$. Since the order parameter $\Delta(x)$ is periodic with the period $2\pi L$, 
it reduces the translation-invariance but still has periodic eigenfunctions. 
The equation is satisfied for all $x$ when $\gamma_2=\gamma_1+i/L$  and
\beq
\label{coefficients1}
\frac{a_2}{a_1}=\frac{E-d(\gamma_1)}{|\Delta|}
,\ \ \
\frac{a_1}{a_2}=\frac{E+d(\gamma_2)}{|\Delta|}
.
\eeq
After dropping the index of $\gamma_1$ (i.e., $\gamma\equiv \gamma_1$)
the operator-valued $\vec{h}$ in Eq. (\ref{su(2)_ham}) becomes a number-valued vector through
\beq
H=d_+(\gamma)\sigma_0+\vec{h}\cdot\vec{\sigma}
\ \ {\rm with}\ \ 
\vec{h}=(\Delta',\Delta'',d_-(\gamma))^T
,
\eeq
$d_\pm=[d(\gamma)\mp d(\gamma+i/L)]/2$ and the $2\times 2$ unit matrix $\sigma_0$.
Eq. (\ref{coefficients1}) implies the quadratic equation in $E$ 
\beq 
\label{eigen_equ1}
[E-d(\gamma)][E+d(\gamma+i/L)]=|\Delta|^2
,
\eeq
which gives the dispersions
\beq
\label{dispersion00}
E_\pm(\gamma)=\frac{d(\gamma)-d(\gamma+i/L)}{2}
\pm\frac{1}{2}\sqrt{[d(\gamma)+d(\gamma+i/L)]^2+4|\Delta|^2}
\eeq
as the solutions. It is crucial that  the values of $\gamma$ are restricted such that 
$E_\pm(\gamma)$ is real,
since the Hamiltonian is Hermitian. Moreover, for a uniform phase, i.e. for $L\to\infty$, this dispersion reduces 
to $E_\pm=\pm\sqrt{d(\gamma)^2+|\Delta|^2}$. 
The corresponding eigenfunctions are obtained from the spinor in Eq. (\ref{spinor0}) as
\beq
\label{spinor1}
\Psi_E(x)
=a_1\pmatrix{
e^{-i x/2L}\cr
\frac{|E-d(\gamma)|}{|\Delta|}e^{ix/2L+i\eta}\cr
}e^{(\gamma +i/2L)x}
,
\eeq
where $\eta$ is the phase of $a_2/a_1$.
The appearance of the conjugate phase factors $e^{\pm ix/2L}$ in the two spinor components is well known 
from the theory of superconductors~\cite{nambu60,caroli64}. They exist for the bulk as well as for the edge modes.
It is important to note that a solution with ${\rm Re}(\gamma)\ne 0$ decays or 
grows exponentially with $x$. Therefore, such solutions can only exist on sample edges. 
In other words, if the system has no edges (e.g. a torus) only solutions with imaginary 
$\gamma$ are allowed.

\begin{figure}[t]
\begin{center}
\includegraphics[width=0.7\linewidth]{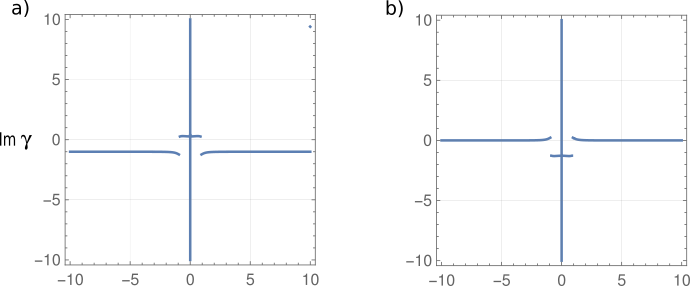}\\
\includegraphics[width=0.7\linewidth]{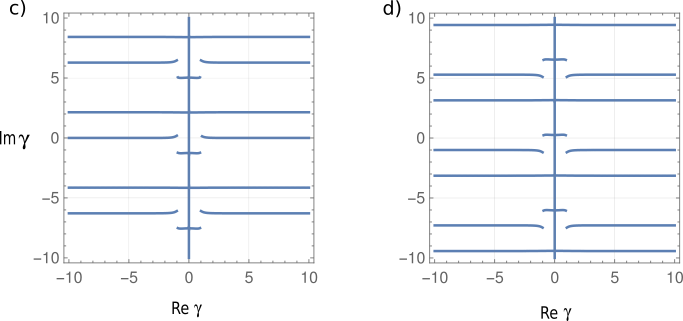}
\caption{
Real-E spectrum: 
The blue curves present values of the complex wave number $\gamma$, where $E_\pm(\gamma)$ is real.
In this example the model parameters are $|\Delta|=L=1$. 
Continuum model: a) $E_+(\gamma)$ and b) $E_-(\gamma)$. Tight-binding model: 
c) $E_+(\gamma)$ and d) $E_-(\gamma)$. The vertical lines with ${\rm Re}\gamma=0$ represent the 
plane-wave solutions, while the other curves represent exponential solutions. 
The actual value of $E_\pm(\gamma)$ is not indicated here but can be calculated separately:
In Fig. \ref{fig:3}b some real energies for $E_+(\gamma)$ of the tight-binding Hamiltonian are presented. 
Examples for $E_\pm(\gamma)$ along the vertical line in b) and c) are plotted in Fig. \ref{fig:4}.
 }
\label{fig:2}
\end{center}
\end{figure} 

\subsection{Bloch vector and winding number}
\label{sect:bloch_vector}

Next we analyze the winding number of the spinor. 
First, we note that the spinor in Eq. (\ref{spinor1}) without the exponential factor is parametrized 
by the coordinate $x$ and it is invariant under $x\to x+4n\pi L$ for an integer $n$.
The winding number is observable and usually related to the Berry phase, where the latter 
can be calculated from the Berry connection~\cite{berry84,simon83}. Alternatively,
the spinor can also be associated with the three-dimensional Bloch vector
\beq
\label{bloch_v00}
\vec{s}(x)=\frac{\Psi(x)\cdot\vec{\sigma}\Psi(x)}{\Psi(x)\cdot\Psi(x)}
,
\eeq
which is the expectation value of the Pauli vector $\vec{\sigma}$. In Maxwell theory this vector is known
as the Stokes vector for the characterization of the electromagnetic field 
and provides the Pancharatnam phase~\cite{pancharatnam56,berry87,avishai25}. It is 
directly linked to the Berry curvature of the electromagnetic field~\cite{ziegler18}.
In the present case
the winding of the spinor can be associated with the three-dimensional Bloch vector of 
Eq. (\ref{bloch_v00}). Its components  characterize the algebraic
relations between the two spinor components of the solution $\Psi_E(x)$ of Eq. (\ref{spinor1}):
\beq
\label{bloch_v0}
s_1=\frac{2b}{1+b^2}\cos(x/L+\eta) 
, \ \
s_2=\frac{2b}{1+b^2}\sin(x/L+\eta) 
, \ \
s_3=\frac{1-b^2}{1+b^2}
,
\eeq
where $b=|a_2/a_1|=|E-d(\gamma)|/|\Delta|$. 
Interestingly, the $x$ dependence of the Bloch vector for bulk and for edge modes differs  
by the phase shift $\eta$ and by different values of $b$, since both
depend on $\gamma$. We will return to this behavior in Sect. \ref{sect:edges}.
The Bloch vector is invariant under $x\to x+2n\pi L$ for an integer $n$, which is half of the 
periodicity of the spinor. A
closed trajectory along the $x$-direction on a torus or ring of length $l$ maps onto
the Bloch sphere as a trajectory with constant latitude that is determined by $s_3$
and with the winding number $w=l/2\pi L$. Thus, the winding number depends only on
the order parameter phase, while the radius of the trajectory $2b/(1+b^2)$ depends on
$\gamma$ and $E$. Boundary conditions restrict the parameter $L$.
Here we assume periodic boundary conditions, such that we get 
$L=l/2\pi n$ with integer $n$, which implies $w=n$.
This means that we can determine the integer winding number by choosing the phase
of the order parameter $\varphi=x/L$. For $E=d(\gamma)$, where $b=0$, the Bloch vector is on the
north pole of the Bloch sphere, while for $b=1$ it is on the equator, and moves 
to the southern hemisphere for $b>1$.

Only the first two components of $\vec{s}$ depend on the phase of $\Delta$, the third component is real
because it is the expectation value of the diagonal matrix $\sigma_3$.
Since the winding number is determined by the phase, for a fixed $|\Delta|$ we can calculate
it using the $s_1$ - $s_2$ - projected trajectory. 
In general, we can express the winding number of the Bloch vector with respect to the north-south axis of the
Bloch sphere as the integral
\beq
\label{invariant}
w=
\frac{{\rm sgn}(s_3)}{2\pi}\int_0^{2\pi}\vec{e}_3\cdot\left[\vec{s}\times
\frac{d\vec{s}}{d\alpha}\right]\frac{1}{s_1^2+s_2^2}d\alpha
={\rm sgn}(s_3)\frac{\epsilon_{3jk}}{2\pi }\int_0^{2\pi}\frac{s_j}{s_1^2+s_2^2}
\frac{ds_k}{d\alpha}d\alpha
\eeq
with $\alpha=x/R$ for the loop of radius $R$ in Fig. \ref{fig:1}.
The unit vector $\vec{e}_3$ defines the north-south axis of the Bloch sphere 
as well as its orientation by ${\rm sgn}(s_3)$, and $\epsilon_{ijk}$  is the Levi-Civita tensor. 
As explained in App. \ref{app:winding}, this integral links the Bloch vector directly with the 
winding number $w$. 

Since the Bloch vector is observable, it
provides a quantity to access and characterize the winding number of  the quasiparticle wave function experimentally.
The Hamiltonian $\vec{h}\cdot\vec{\sigma}$ can be formally understood as the energy of a pseudospin $\vec{\sigma}$
that is coupled to a pseudomagnetic field $\vec{h}$. The real and the imaginary part of the order parameter
are two components of the pseudomagnetic field. In this picture the Bloch vector in Eq. (\ref{bloch_v00}) is
the local expectation value of the pseudospin with respect to the spinor wave function at a given energy. As we go along the 
$x$-direction of the superconducting ring in Fig. \ref{fig:1}, the pseudomagnetic field rotates according to Fig. \ref{fig:5}a,
while the Bloch vector rotates on the Bloch sphere at a fixed latitude according to Fig. \ref{fig:5}b; i.e., the third component of 
the pseudospin is fixed as $s_3=(1-b^2)/(1+b^2)$ with $b=|E-d(\gamma)|/|\Delta|$. Thus the Bloch vector represents 
a pseudospin ordering in the superconductor, which is determined by the energy $E$, the magnitude of the order 
parameter $|\Delta|$ and by the eigenvalue $d(\gamma)$ of the kinetic operator $D$. For $b=0$ there is
ferromagnetic order, for $b=1$ there is helical order and for $0<b<1$ a combination of ferromagnetic and 
helical order. The phase shift $\eta$ in Eq. (\ref{bloch_v0}), which is $0$ in the upper band and $\pi$ 
in the lower band of bulk modes, defines a phase shift of the pseudospin relative to the order parameter phase. 
The value of $\eta$ varies continuously with energy between $\eta=0$ and $\eta=\pi$ for the edge modes, 
a property that distinguishes between bulk and edge modes~\cite{ziegler25b}. 
This will be discussed for the BdG Hamiltonian in Sect. \ref{sect:edges}

Finally, we note that a gauge transformation for superconductors and superfluids connects the kinetic term with
the phase of the order parameter~\cite{nambu60,byers61,caroli64}. In our specific case this gauge
transformation reads as the unitary transformation
\beq
\label{unitary_tr0}
\vec{h}\cdot\vec{\sigma}\to \pmatrix{
e^{i\varphi/2} & 0 \cr
0 & e^{-i\varphi/2} \cr
}
\pmatrix{
D & |\Delta|  \cr
|\Delta| & -D \cr
}\pmatrix{
e^{-i\varphi/2} & 0 \cr
0 & e^{i\varphi/2} \cr
}
=\pmatrix{
\tilde{D} & |\Delta| e^{i\varphi} \cr
|\Delta|e^{-i\varphi} & -\tilde{D}^* \cr
}
,\ \ 
\tilde{D}=e^{i\varphi/2}De^{-i\varphi/2}
,
\eeq
which creates the phase factor $e^{i\varphi}$ for the real order parameter $|\Delta|$ by a simultaneous 
transformation of the operator $D\to\tilde{D}$.
This reflects the fact that the phase factor of the vortex-like order parameter can be represented as a vector 
potential in the kinetic operator $D$, for instance, by an external magnetic field outside of the 
superconductor~\cite{byers61}. The transformation implies that the pseudospin ordering can also be created
by a vector potential. More details will be discussed subsequently for specific Hamiltonians.

Many physical quantities depend only on the spectral properties of the Hamiltonian.
For instance, trace expressions such as the thermodynamic quantities 
or the density of states. This is also the case for expressions represented by an inner
product. These quantities are invariant under unitary transformations
including the transformation of the operator domain. Correlation functions,
on the other hand, may not be invariant under general unitary transformations, such as
the relation of the spinor components in Eqs. (\ref{spinor0}) or (\ref{spinor1}),
whose real and imaginary parts are observable. 

\begin{figure}[t]
\begin{center}
a)
\includegraphics[width=0.25\linewidth]{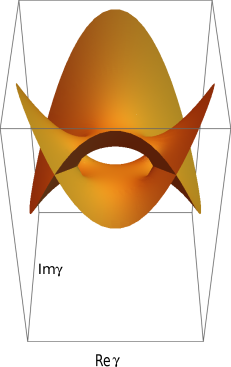}
b)
\includegraphics[width=0.6\linewidth]{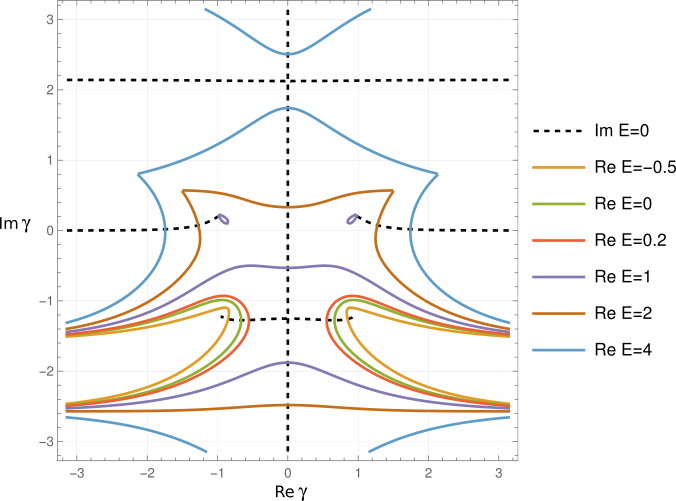}
\caption{
a) Riemann surface of the real part ${\rm Re}E$ of the complex energy $E$ of the continuous
BdG Hamiltonian in Eq. (\ref{dispersion_c}) with $|\Delta|=L=1$. 
It is symmetric with respect to ${\rm Re}\gamma$.
b) Contour plot of the complex tight-binding BdG spectrum: The black dashed contours indicate ${\rm Im} E=0$, 
the colored contours represent constant ${\rm Re}E$. The curve crossings represent real eigenvalues, whose 
values are color encoded.
}
\label{fig:3}
\end{center}
\end{figure} 

\section{Bogoliubov de Gennes Hamiltonians}
\label{sect:examples}

In the following we will consider two examples for the operator $D$, namely the BdG
Hamiltonian in a continuous space and its tight-binding version on a square lattice. 
The BdG Hamiltonian with different symmetries have been intensively studied for translation-invariant 
systems ~\cite{nambu60,caroli64,skurativska20,ono20,geier20}. We will rely here on its simplest form but break the
translational invariance by a periodic order parameter $\Delta(x)$, as introduced in the previous
section. 
The existence of a spatially modulated order parameter with a characteristic wave 
vector  $\bq$ is known from the Fulde-Ferrell (FF) phase~\cite{fulde64} or as the 
Larkin-Ovchinnikov (LO) phase~\cite{larkin64} of a 
superconductor in a magnetic field. $\bq$ is determined by the 
linearized gap equation and depends on the external magnetic field, the temperature and
the Fermi surface~\cite{shimahara09}. 
A simpler realization is the superconducting ring
of Fig. \ref{fig:1}, where the order-parameter oscillations are induced by a supercurrent.
In the following we will focus on the FF phase, where $|\Delta|$ is uniform.

\subsection{Continuous BdG Hamiltonian}
\label{sect:cont_bdg}

We choose the $SU(2)$ Hamiltonian $\vec{h}_{BdG}\cdot\vec{\sigma}$ with 
$D=-\partial_x^2-\partial_y^2$ acting on a finite continuous space:
\beq
\label{BdG01}
\vec{h}_{BdG}=(\Delta',\Delta'',-\partial_x^2-\partial_y^2)
.
\eeq
When we assume that $\Delta$ is uniform in the $y$ direction, we can apply
the Fourier transformation $-\partial_y^2\to k_y^2$.
The corresponding eigenvalue problem of the quasiparticles with energy $E$ 
reads as the BdG equation
\beq
\label{BdG1}
\pmatrix{
-\partial_x^2 +k_y^2 & |\Delta|e^{ix/L} \cr
|\Delta|e^{-ix/L} & \partial_x^2 -k_y^2\cr
}\pmatrix{
\psi_1\cr
\psi_2\cr
}=E\pmatrix{
\psi_1\cr
\psi_2\cr
}
,
\eeq
whose solutions describe the wavefunctions along a torus or a cylinder in the $x$
direction, respectively. 
The spatial variation of the order parameter phase $\varphi$ induces a supercurrent ${\bf j}_s$
according to the Ginzburg-Landau supercurrent-phase relation~\cite{tinkham96}
\beq
\label{CPR}
{\bf j}_{\rm s}=\frac{\hbar e^*}{m^*}|\Delta|^2(\nabla\varphi-\frac{e^*}{\hbar c}{\bf A})
\eeq
in the presence of a vector potential ${\bf A}$. $e^*$ and $m^*$ are the charge and the mass of
the Cooper pairs. This relation provides the unitary transformation in Eq. (\ref{unitary_tr0}) by 
an appropriate choice of the supercurrent and the vector potential. In particular, the special choice
$\varphi=x/L$ creates the phase of Eq. (\ref{BdG1}). This means that the winding number $w$
of the Bloch vector, defined by the integral in Eq. (\ref{invariant}), is determined by the supercurrent
through the Ginzburg-Landau relation. Moreover, we note that the quasiparticle currents depend 
on the winding number $w$. They were studied explicitly, for instance, in Ref.~\cite{ziegler25a}. 

Neglecting the vector potential in Eq. (\ref{CPR}), the phase gradient $\nabla\varphi$ is proportional to
the supercurrent. This suggests that we obtain $\varphi=x/L$ when we create a constant supercurrent 
in $x$--direction with the help of an external field.
An experimental realization would be a superconducting ring, as illustrated in Fig. \ref{fig:1}, which is reminiscent 
of a macroscopic vortex. The resulting phase jump is not a problem as long as the complex order parameter 
$\Delta(x)$ is continuous, since the phase is not observable due to the $U(1)$ symmetry of the superconductor.
A continuous $\Delta(x)$ implies a quantized phase jump $2\pi n$ ($n=\pm1,\pm2,\ldots$).

\begin{figure}[t]
\begin{center}
a)
\includegraphics[width=0.3\linewidth]{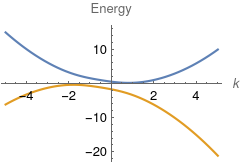}
b)
\includegraphics[width=0.3\linewidth]{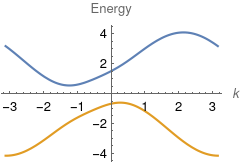}
\caption{
Energy dispersion $E_\pm(k)$ of the plane-wave solution for a) the continuous BdG Hamiltonian 
and b) for the tight-binding BdG Hamiltonian with $|\Delta|=L=1$.
These dispersions correspond to the vertical lines in Fig. \ref{fig:2}.
}
\label{fig:4}
\end{center}
\end{figure} 

From Sect. \ref{sect:model} we get for this special case the $\gamma\to-\gamma$ symmetric
eigenvalues $d(\gamma)=-\gamma^2+k_y^2$ and Eq. (\ref{eigen_equ1}) for the energy eigenvalues becomes
\beq
\label{quartic_eq}
[E+\gamma^2-k_y^2][E-(\gamma+i/L)^2+k_y^2]=|\Delta|^2
.
\eeq
Moreover, considering a constant solution in $y$ direction (i.e., $k_y=0$), we obtain
as the solution of the quadratic equation $(E+\gamma^2)[E-(\gamma+i/L)^2]=|\Delta|^2$ 
for $E$ the dispersion
\beq
\label{dispersion_c}
E_\pm(\gamma)
=\frac{i\gamma}{L}-\frac{1}{2L^2}
\pm\frac{1}{2}\sqrt{[\gamma^2+(\gamma+i/L)^2]^2+4|\Delta|^2}
.
\eeq
This is not symmetric under $\gamma\to-\gamma$ but its real part is symmetric with respect
to ${\rm Re}\gamma\to-{\rm Re}\gamma$ (cf. Fig. \ref{fig:3}).
With $\gamma=\kappa +ik$, where $k$ and $\kappa$ are real, 
we can distinguish plane-wave solutions for $\kappa=0$ and exponential solutions for $\kappa\ne0$.
Exponential means here that the absolute value of the
solution either decreases or increases exponentially with $x$.
Employing the plane-wave solution $\gamma=ik$ we obtain
$(E-k^2)[E+(k+1/L)^2]=|\Delta|^2$. This gives for the dispersion in Eq. (\ref{dispersion_c})  
\beq
\label{energy_levels}
E_\pm(ik)
=-\frac{k}{L}-\frac{1}{2L^2}\pm\frac{1}{2}\sqrt{
[k^2+(k+1/L)^2]^2+4|\Delta|^2}
,
\eeq
which is real for any $k$ with $-\infty<k<\infty$.
The $1/L$ terms can be understood as the Doppler shift caused by the gauge transformation
in Eq. (\ref{unitary_tr0}) .
This dispersion is plotted in Fig. \ref{fig:4}a) for $|\Delta|=L=1$.
Expansion for small $k$ yields a gap and a linear $k$ term,
while the asymptotic behavior for large $k$ is parabolic:
\beq
E_\pm\sim \pm k^2
.
\eeq
For the zero mode $E=0$ and $k_y=0$ we get directly from Eq. (\ref{quartic_eq}) the quartic equation
$\gamma^2(\gamma+i/L)^2=-|\Delta|^2$ that reduces to the two quadratic 
equations $\gamma(\gamma+i/L)=\pm i|\Delta|$ with the four solutions
\beq
\label{zero_modes}
\gamma=-\frac{i}{2L}\pm i\sqrt{i|\Delta|+1/4L^2}
\ \ {\rm and}\ \ 
\gamma'=-\frac{i}{2L}\pm i\sqrt{-i|\Delta|+1/4L^2}
.
\eeq
These solutions are exponential for $|\Delta|>0$, since we have a gap for the plane-wave solution. 
There are other exponential solutions for real $E\ne0$,
as indicated by the horizontal blue curves on the complex $\gamma$ plane in Figs. \ref{fig:2}a, \ref{fig:2}b. 
The corresponding values of $E$ along these curves are not indicated in these two-dimensional
plots, but are plotted separately in Fig. \ref{fig:3}b for some characteristic values of $\gamma$.

The unitary transformation of Eq. (\ref{unitary_tr0}) implies in this case
\beq
\label{gauge_tr1}
-\partial_x^2\to(i\partial_x+1/2L)^2
,
\eeq
where $1/2L$ represents the gradient of a gauge field. 
This reflects the relation in Eq. (\ref{CPR}) in terms of the quasiparticle Hamiltonian, 
in which an additional gauge field changes the gradient of the 
order-parameter phase.

Finally, for the plane-wave solutions we have $b=|E-k^2|/|\Delta|$ in Eq. (\ref{bloch_v0}). 
The winding number for the ring-like geometry in Fig. \ref{fig:1} is determined by the integral in Eq.
(\ref{invariant}): The integration is parametrized in this case with $\alpha=x/R$ along the ring's
circumference $2\pi R$, where $R$ is the radius of the ring. Periodic boundary conditions
on the ring require $R/L=n$ with $n=0,\pm1,\ldots$, and yield the winding number $w=n$.
This means that the winding numbers in the upper and in the lower band are the same. 

\subsection{Tight-binding BdG Hamiltonian on a square lattice}
\label{sect:bdg_tb}

In general, the spectral properties of a system in the continuum and on a lattice are quite different.
We will study next whether this is also the case for the Bloch vector of the BdG Hamiltonian.
To this end we consider the case in which $D$ is a difference operator with 
$D\psi_x=\psi_{x+1}+\psi_{x-1}-2\psi_x$ that acts on a discrete lattice with unit lattice spacing.
Here we have only considered the $x$ direction, the $y$ direction is analogue.
Its eigenvalue condition reads in this case
\beq
De^{\gamma_j x}=e^{\gamma_j(x+1)}+e^{\gamma_j(x-1)}-2e^{\gamma_j x}
=2(\cosh \gamma_j-1)e^{\gamma_j x}
,
\eeq
such that $d(\gamma)=2(\cosh \gamma-1)$ is symmetric with respect to $\gamma\to-\gamma$
and gives with Eq. (\ref{dispersion00}) the dispersion
\beq
\label{dispersion_tb0}
E_\pm(\gamma)
=\cosh(\gamma)-\cosh(\gamma+i/L)
\pm \sqrt{[\cosh(\gamma)+\cosh(\gamma+i/L)-2]^2+|\Delta|^2}
,
\eeq
which is complex for general $\gamma$ and not symmetric.
For $\gamma=ik$ the plane-wave dispersion follows as 
\beq
\label{dispersion_tb}
E_\pm(ik)
=\cos k -\cos(k+1/L)
\pm\sqrt{[\cos(k)+\cos(k+1/L)-2]^2+|\Delta|^2}
,
\eeq
which is real again for any $k$ with $-\pi\le k<\pi$. The dispersion $E_\pm(ik)$
is visualized in Fig. \ref{fig:4}b). 
There are also exponential solutions with a real dispersion. They have nonzero ${\rm Re}\gamma$ values
and are indicated as horizontal curves in Figs. \ref{fig:2}c and \ref{fig:2}d.  

Inserting the eigenvalues $d(\gamma)$ into the spinor of Eq. (\ref{spinor1}) and into the Bloch 
vector of Eq. (\ref{bloch_v0}) enables us to calculate the winding number for the tight-binding
BdG Hamiltonian. 
For plane-wave solutions $b=|a_2/a_1|=|E-d(\gamma)|/|\Delta|=|E-2(\cos k-1)|/|\Delta|$.
Thus, the Bloch vector is affected by the introduction of a lattice. However, since $b$ does
not depend on $x$, it drops out of the integrand in Eq. (\ref{invariant}), and we
get the same winding number as in the continuum case.
Finally, the unitary transformation of Eq. (\ref{unitary_tr0}) reads in this case
\beq
\tilde{D}\psi_x=e^{-i/2L}\psi_{x+1}+e^{i/2L}\psi_{x-1}-2\psi_x
,
\eeq
where $1/2L$ represents a Peierls phase on the hopping elements in $x$ direction, in analogy
with the gauge transformation of the continuous BdG Hamiltonian in Eq. (\ref{gauge_tr1}).

\section{Exponential solutions}
\label{sect:edges}

Now we turn to a system with an edge. As a specific example we consider the superconducting
ring of Fig. \ref{fig:1} and add a cut or a weak link in radial direction at $x=0$. 
While $|\Delta|$ is fixed on the ring, the cut at $x=0$ may jump.
Thus, the order parameter of this system does not have a quantized phase but 
exhibits a phase slip~\cite{petkovic16} at the cut. 
This will create quasiparticle edge modes with an exponential behavior in $x$ direction
and energies inside the gap. As mentioned before, a characterization of the edge mode is given
by the phase shift $\eta$ in Eq. (\ref{bloch_v0}). Since $\gamma$ is complex for edge modes, the
phase ${\rm arg}[E-d(\gamma)]$ can be non-zero. For instance, in the limit $L\to\infty$ we
get from Eq. (\ref{quartic_eq}) two eigenvalues of $D$ and $E-d(\gamma_\pm)=E\mp\sqrt{E^2-|\Delta|^2}$.
This implies $\eta={\rm arg}(E-d(\gamma_\pm))=0$ for the upper bulk band and 
$\eta={\rm arg}(E-d(\gamma_\pm))=\pi$ for the lower bulk band. For the energies of the edge modes 
$-|\Delta|<E<|\Delta|$ we get the phase shift
\beq
\label{phase1}
\eta=\cases{
\pi-\arctan(\sqrt{|\Delta|^2/E^2-1})  & $-|\Delta|<E\le 0$ \cr
\arctan(\sqrt{|\Delta|^2/E^2-1}) & $0\le E<|\Delta|$ \cr
}
,
\eeq
which describes a continuous rotation of the pseudospin from $\pi$ to $0$ during an energy
change from $E=-|\Delta|$ to $E=|\Delta|$. Finite $L$ can be treated in a $1/L$ expansion and gives
small corrections to $\eta$ in Eq. (\ref{phase1}) of order $1/L$.
To find the exponential eigenfunctions inside the gap, we can either solve Eq. (\ref{eigen_equ1})
directly or employ the analytic bulk-edge connection~\cite{ziegler25a,ziegler25b}. 
The eigenvalues $d(\gamma)=-\gamma^2+k_y^2$ of the continuous BdG Hamiltonian and
$d(\gamma)=2(\cosh \gamma-1)+2(\cos k_y-1)$ of the tight-binding BdG Hamiltonian are both real 
for the bulk states. This suggests that the analytic continuation starts from a plane-wave (bulk) solution. 
Then the analytic continuation $ik\to\gamma$ of the real wave number $k$ into the complex 
plane is applied. In general, this leads to a complex energy, which violates the eigenvalue 
equation for real energies. Therefore, the additional condition of a real $E_\pm(\gamma)$ 
must be enforced. The results of the analytic bulk-edge connection are visualized in 
Figs. {\ref{fig:2}a --{\ref{fig:2}d: 
For the continuous BdG Hamiltonian we obtain the real-E curves in Figs. \ref{fig:2}a, {\ref{fig:2}b, while
the tight-binding BdG Hamiltonian gives Figs. \ref{fig:2}c, {\ref{fig:2}d. In both cases the vertical lines at 
${\rm Re}\gamma=0$ represent the plane-wave solutions, while the other curves represent exponential 
solutions. The periodic behavior of the tight-binding spectrum in Eq. (\ref{dispersion_tb}) for imaginary 
$\gamma$ is reflected by the repeated real-E curves in Figs. \ref{fig:2}c, {\ref{fig:2}d. Apart from this feature, 
the real-E spectrum is similar for both Hamiltonians and their real-E curves are symmetric with 
respect to ${\rm Re}\gamma\to-{\rm Re}\gamma$.

The Bloch vector in Eq. (\ref{bloch_v0}) is valid for plane-wave as well as for exponential solutions 
of Eq. (\ref{BdG1}), an exponential solution might have a phase shift $\eta\ne0$ and a different 
$b$ though. The winding number is directly linked to the phase of the order parameter
$x/L$. In general, we might consider non-periodic boundary conditions, for instance, in a
superconducting ring with a Josephson barrier. This indicates that the winding number of the bulk 
modes differ from those of the edge modes simply due to different boundary conditions.
The latter depend on the specific set-up of the physical system.
It is beyond the scope of the present work to elaborate on this experiment-specific issue.
In any case, though, the boundary conditions must also lead to a self-adjoint
Hamiltonian, which requires that the energy eigenstates for different eigenvalues are 
orthogonal. 
There are four solutions for $\gamma$ of the quartic Eq. (\ref{quartic_eq}) at a fixed
energy $E$. In particular, for $E=0$ we get from Eq. (\ref{zero_modes})
with $c'={\rm Re}( i\sqrt{i|\Delta|+1/4L^2})$ and $c''={\rm Im} (i\sqrt{i|\Delta|+1/4L^2})$
for the $x$-dependent phase factor in Eq. (\ref{spinor1})
\beq
\psi(x)=e^{-c'x}(a_{++}e^{ic''x}+a_{--}e^{-ic''x})+e^{c'x}(a_{-+}e^{ic''x}+a_{+-}e^{-ic''x})
.
\eeq
To get $\psi(L)=0$ we choose the coefficients such that the factors inside the brackets vanish at the
boundary $x=L$
\beq
a_{++}=-e^{-2ic'' L}a_{--}
\ ,\ \
a_{-+}=-e^{-2ic'' L}a_{+-}
.
\eeq
Then we only need $a_{--}=-a_{+-}$ to get also a vanishing function at $x=0$ 
\beq
\psi(0)=(a_{--}+a_{+-})(1-e^{-2ic'' L})=0
.
\eeq
This choice corresponds to Dirichlet boundary conditions, which guarantee that the Hamiltonian is self-adjoint.
Dirichlet boundary conditions can also be applied in a similar way to edge modes with $E\ne0$.

\begin{figure}[t]
\begin{center}
\includegraphics[width=0.5\linewidth]{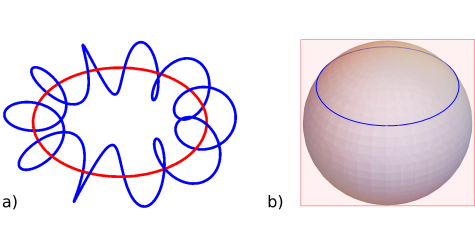}\\
\includegraphics[width=0.5\linewidth]{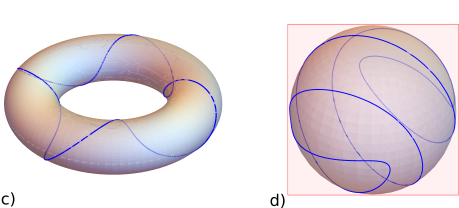}
\caption{
Top row is for the BdG Hamiltonian: a) A superconducting order parameter 
$(\Delta',\Delta'')=(\cos(10x),\sin(10x))$ with $0\le x<2\pi$
(blue) along a superconducting ring (red) implies b) a map on the Bloch sphere. The winding
number of this trajectory is $w=10$; i.e., the trajectory winds 10 times around the Bloch sphere.
The latitude on the Bloch sphere depends on the parameter $b$, which is $b=1/2$ in this example.\\ 
Bottom row is for the $\pi$-flux Hamiltonian:
c) a closed helical trajectory on a torus with $n_x=4$, $n_y=1$ windings and d)
its map on the Bloch sphere. The winding number of this trajectory with respect to the north-south axis 
of the Bloch sphere is $w=0$.
}
\label{fig:5}
\end{center}
\end{figure} 

\section{Discussion and summary}
\label{sect:discussion}

The mapping $\vec{h}\cdot\vec{\sigma}\to\Psi\to\vec{s}$ of Sect. \ref{sect:model}, where 
$\vec{h}\cdot\vec{\sigma}$ is the Hamiltonian, $\Psi$ is an eigenspinor of $\vec{h}\cdot\vec{\sigma}$
and $\vec{s}$ is the three-dimensional
Bloch vector, provides a triple of relevant quantities to characterize an $SU(2)$--based quantum system.
The Bloch vector $\vec{s}$ is an observable that visualizes the winding number of the quantum
state.
To study the effect of the order-parameter phase on the quasiparticles of a superconductor, we have 
introduced  the periodic order parameter $\Delta=|\Delta|e^{ix/L}$
with the parameter $L$, which induces the spatial winding number $w$ of the Bloch vector.
$L$ depends on the boundary conditions and can be controlled in the case of the BdG Hamiltonian 
through the supercurrent as well as through the vector potential of an external electromagnetic field.
The appearance of a periodic order-parameter phase can be linked with
the picture of a macroscopic vortex. According to our discussion in Sect. \ref{sect:examples}, the
order parameter phase determines the winding number in the case of periodic boundary conditions.
In Figs. \ref{fig:5}a, b an example with the order-parameter phase $\varphi=10x/2\pi$ 
and its corresponding trajectory on the Bloch sphere are depicted.

An important question regarding the observation of the winding number by using the
Bloch vector is whether a closed Bloch-vector trajectory is always accompanied by closed spinor
trajectory? The answer is no because this depends on the definition of the spinor.
Our specific spinor definition in Eq. (\ref{spinor1}) indicates that the winding numbers do not agree:
The spinor of Eq. (\ref{spinor1}) has the periodicity $4n\pi L$ but the
Bloch vector in Eq. (\ref{bloch_v0}) is periodic in $x$ with period $2n\pi L$. 

In the BdG Hamiltonian the winding number $w$ of the Bloch vector is induced by the phase
of the order parameter. This implies that a real order parameter gives $w=0$. But there are other 
mechanisms to create a Bloch vector with $w\ne0$ for $SU(2)$ Hamiltonians, for instance, in chiral systems.
We briefly mention here the $\pi$-flux Hamiltonian~\cite{fradkin} as an example with 
\beq
\vec{h}=(\sin k_x, \sin k_y,m\cos k_x\cos k_y)^T
\eeq
instead of Eq. (\ref{su(2)_ham}), where
$(k_x,k_y)$ is a wave vector on a torus. It describes a particle hopping on a square lattice with flux $\pi$ per
plaquette. We consider a closed helical trajectory on the torus that winds $n_x$ times around 
the small circle and $n_y$ around the large circle.
An example with $n_x=4$ and $n_y=1$ is visualized in Fig. \ref{fig:5}c.  This trajectory, 
which can be understood as an edge on the torus,
is mapped onto the Bloch sphere, as demonstrated in Fig. \ref{fig:5}d, results in a twisted loop on 
the Bloch sphere whose winding 
number is $w=0$ according to Eq. (\ref{invariant}). This is also the case for other values of $n_x\ge 1$.
In contrast to the rotation of the order-parameter phase in the BdG Hamiltonian, an increasing $n_x$
on the torus does not result in an increasing $w$ for the $\pi$-flux Hamiltonian. On the other hand, 
a non-helical loop on the torus, e.g., one with $n_x=n_y=0$, has a winding number $w=1$ or $w=-1$, 
depending on the position of the loop center on the torus. Thus, the winding number in the $\pi$-flux 
Hamiltonian is robust with $w=0, \pm1$, only depending on the trajectory on the torus.
The Bloch vector, written as the expectation value of the pseudospin,
is invariant under a unitary transformation, which is seen by
direct inspection with the transformation of Eq. (\ref{unitary_tr0}).
Then Eq. (\ref{invariant}) is invariant and implies for the winding number 
\beq
\label{winding2}
w=\frac{\varphi(2\pi)-\varphi(0)}{2\pi}
,
\eeq
where the parametrization $\varphi(\alpha)$ of the phase with $0\le \alpha<2\pi$ has been used.
This phase depends on the specific model we consider and the special closed trajectory. For the BdG 
Hamiltonian on a superconducting
ring it depends on the superconducting order parameter, while for the $\pi$-flux Hamiltonian it is 
determined by the trajectory
on the torus of the underlying physical space. An example is a small loop for the $\pi$-flux Hamiltonian, 
which gives $\varphi(\alpha)=\alpha$ for a loop centered at $k_x=k_y=0$ or 
$\varphi(\alpha)= -\alpha$ for a loop centered at $k_x=0$, $k_y=\pi$, leading to $w=\pm1$
according to Eq. (\ref{winding2}). This case is related to the corresponding Chern number, which
effectively calculates a simple closed loop without winding around the torus through the Berry curvature~\cite{fradkin}.

Focusing on the BdG Hamiltonian, a giant vortex can be created on a superconducting 
disk~\cite{bruyndoncx99,kanda04} or on a superconducting ring~\cite{liu21}, which provides the 
modulated order-parameter phase. In such a system it should be possible to
study the properties of the quasiparticles as a function of the vorticity. 
As an observable, the Bloch vector is experimentally accessible through linear response with respect
to the order parameter. 
In terms of the pseudospin ordering, discussed in Sect.  \ref{sect:model}, the Bloch vector field can be probed 
by optical spectroscopy with polarized light or within angle-resolved photoemission spectroscopy 
(ARPES)~\cite{schueler22,yam25}.
These are established methods to analyze pseudospin properties in graphene-like two-band materials. Tunneling
spectroscopy~\cite{anthore03} is also an option because the local density of states is sensitive to the pseudospin.
In particular, we get from a change of the energy eigenvalues the first two 
components of the Bloch vector as $s_1=\partial_{\Delta'}E$ and $s_2=\partial_{\Delta''}E$, 
which are directly linked to the group velocity~\cite{liu25}. 
They are relevant for the winding number, and we can insert them directly into Eq. (\ref{invariant}).
The single-ring geometry can be extended to a network of superconducting links as a host of several 
vortices. This system represents a more complex order parameter for topologically protected quasiparticles,
which could be useful for robust quantum-technological applications. 

Finally, it should be emphasized that the Hamiltonians discussed in this work are Hermitian.
An extension by including non-Hermitian terms, as used in the concept of ``non-Bloch BdG
Hamiltonians''~\cite{yao18,yokomiz021}, might be interesting but should be left for future projects.
In particular, to compare the role of edge modes in the Hermitian case with their role in the non-Hermitian 
case could offer new insights into the effect of the environment on the quantum system.
Another extension of the present work could be based on other order parameters.
For instance, we can consider piecewise linear phases on short intervals with alternating
slopes $\varphi=\pm x/L$, where the phase is still continuous. The matching of the
eigenfunctions at the sign-switching points creates additional exponential solutions,
creating a special kind of localized wave function. Moreover, if the steps between sign switches
are random, this mimics a random phase similar to those in disordered systems.

\appendix

\section{Winding number as an invariant on the Bloch sphere}
\label{app:winding}

We consider a Bloch sphere whose north-south axis is parallel to the unit vector $\vec{e}_3$
and the Pauli matrix $\sigma_3$ is the $2\times2$ diagonal matrix.
The Bloch vector is parametrized as usual with the two angles $\varphi$, $\theta$ as
\beq
\label{3d_vector}
\vec{s}=\pmatrix{
\cos\varphi\sin\theta \cr
\sin\varphi\sin\theta \cr
\cos\theta\cr
}
.
\eeq
Both angles depend on the parameter $\alpha$ with the same values at $\alpha=0$ and $\alpha=2\pi$.
Then we define the winding around the $3$-axis as the integral
\beq
w:=
\frac{{\rm sgn}(s_3)}{2\pi}\int_0^{2\pi}\vec{e}_3\cdot\left[\vec{s}\times
\frac{d\vec{s}}{d\alpha}\right]\frac{1}{s_1^2+s_2^2}d\alpha
={\rm sgn}(s_3)\frac{\epsilon_{3jk}}{2\pi }\int_0^{2\pi}\frac{s_j}{s_1^2+s_2^2}
\frac{ds_k}{d\alpha}d\alpha
.
\eeq
Here $\sqrt{s_1^2+s_2^2}$ is the normalization of the vector $(s_1,s_2)^T$.
This integral acts only on the $s_1$ - $s_2$ projected Bloch vector, which
is a consequence of the fact that the third vector component $s_3$ is the reference axis for the winding of the 
spinor. Therefore, $s_3$ does not contribute due to the Levi-Civita tensor $\epsilon_{3jk}$.
Moreover, the differential with respect to $\theta$ drops out due to
\beq
s_2\frac{ds_1}{d\theta}=s_1\frac{ds_2}{d\theta}
.
\eeq
Thus, we get from Eq. (\ref{3d_vector})
\beq
w=\frac{{\rm sgn}(s_3)}{2\pi}\int_0^{2\pi}(\cos^2\varphi+\sin^2\varphi)
\frac{d\varphi}{d\alpha}d\alpha
=\frac{{\rm sgn}(s_3)}{2\pi}\int_0^{2\pi}\frac{d\varphi}{d\alpha}d\alpha
={\rm sgn}(s_3)\frac{\varphi(2\pi)-\varphi(0)}{2\pi}
.
\eeq
This is the winding number. We note that $\vec{s}(\alpha)$ can be a multi-valued function on the interval $[0,2\pi)$.
For instance, with $\varphi=n\alpha$ we get for the integral $w=n$. 

It should be noted that there is an ambiguity in the choice of $\vec{e}_3$. Instead of using the south pole as the
reference point of the coordinate system, we could also choose the center of the Bloch sphere. Then
the winding numbers differ by a sign change between the southern and the northern hemisphere.

\section{The $2\times 2$ matrix structure}
\label{app:matrix}

There is an intimate connection between spinors and winding numbers.
As an instructive demonstration of this connection, we consider the $2\times2$ Hamiltonian
\beq
\label{ham_00}
H_0=\pmatrix{
m & z \cr
z^* & -m \cr
}
=\vec{h}_0\cdot\vec{\sigma}
\eeq
with $m$ real , $z=z'-iz''$ and real $z'$, $z''$. The vector $\vec{\sigma}=(\sigma_1,\sigma_2,\sigma_3)$ 
comprises the Pauli matrices and $\vec{h}_0=(z',z'',m)$.
The corresponding eigenvalue problem with eigenvalue $E=\pm\sqrt{m^2+|z|^2}$ reads
\[
\pmatrix{
m & z \cr
z^* & -m \cr
}\pmatrix{
a_1\cr
a_2\cr
}=E\pmatrix{
a_1\cr
a_2\cr
}
,
\]
which separates into two equations
\[
\cases{
a_1=\frac{z}{E-m}a_2 \cr
a_2=\frac{z^*}{E+m}a_1\cr
}
.
\]
A solution of these equations is the spinor
\beq
\label{spinor00}
\Psi=a_1\pmatrix{
1 \cr
z^*/(E+m) \cr
}
=a_1\pmatrix{
1 \cr
(E-m)/z \cr
}
.
\eeq
The related Bloch vector is of unit length and its components read
\beq
\label{bloch_v01}
s_1=\frac{2z'}{{\cal N}_0(E+m)} 
,\ \ 
s_2=\frac{2z''}{{\cal N}_0(E+m)} 
,\ \
s_3=\frac{1}{{\cal N}_0}\left[1-\frac{|z|^2}{(E+m)^2}\right]
\eeq
with the normalization ${\cal N}_0=1+|z|^2/(E+m)^2$. This means that a closed trajectory 
of $z$ on the
complex plane or on some Riemann sheets maps onto a closed trajectory on the Bloch sphere.

\end{document}